# Prediction and Characterization of Two-Dimensional $Zn_2VN_3$

Andrey A. Kistanov,[1,*] Stepan A. Shcherbinin,[2,3] Elena A. Korznikova,[1] Oleg V. Prezhdo[4]

[1] The Laboratory of Metals and Alloys Under Extreme Impacts, Ufa University of Science and Technology, 450076 Ufa, Russia

[2] Peter the Great Saint Petersburg Polytechnical University, 195251 Saint Petersburg, Russia

[3] Institute for Problems in Mechanical Engineering RAS, 199178 Saint Petersburg, Russia

[4] Department of Chemistry, University of Southern California, Los Angeles, California 90089, United States

*Corresponding author: andrei.kistanov.ufa@gmail.com

**Abstract**

A two-dimensional (2D) monolayer of a novel ternary nitride $Zn_2VN_3$ is computationally designed, and its dynamical and thermal stability is demonstrated. A synthesis strategy is proposed based on experimental works on production of ternary nitride thin films, calculations of formation and exfoliation energies, and *ab initio* molecular dynamics simulations. A comprehensive characterization of 2D $Zn_2VN_3$, including investigation of its opto-electronic and mechanical properties, is conducted. It is shown that 2D $Zn_2VN_3$ is a semiconductor with an indirect band gap of 2.75 eV and a high work function of 5.27 eV. Its light absorption covers visible and ultraviolet regions. The band gap of 2D $Zn_2VN_3$ is found to be well tunable by applied strain. At the same time 2D $Zn_2VN_3$ possesses high stability against mechanical loads, point defects, and environmental impacts. Considering the unique properties found for 2D $Zn_2VN_3$, it can be used for application in opto-electronic and straintronic nanodevices.

**TOC**

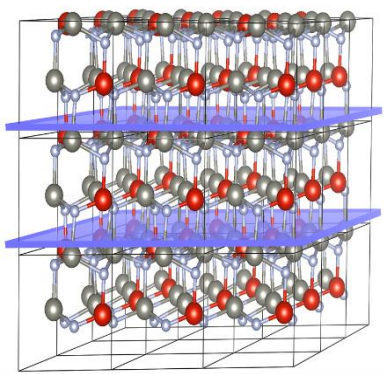

Two-dimensional (2D) materials remain an active field of research in science and engineering over the last 15 years. During that time countless 2D structures with unique superconducting (1), opto-electronic (2,3), magnetic (4), mechanical (5,6), and topological (7) properties have been found. For the most part, the research effort has been directed to the exploration of 2D materials which have bulk counterparts representing anisotropic crystals with layers folded together by van der Waals (vdW) forces (8). Weak vdW interaction between the layers of such 2D materials supports a natural structural separation of the 2D subcells in the crystals via mechanical (9) or liquid-phase (10) exfoliation. Although the deposition and growth technologies are generally available for of 2D materials, control of defects and contamination is not yet compliant with the specifications defined for production (11). High processing temperatures are typically required for high quality 2D materials, complicating their production. Despite these challenges, existing and newly developed 2D materials carry the promise of successful integration into technology and commercial devices during the next decade (11).

Various methods to predict and fabricate novel bulk and 2D materials, including experimental, *ab initio*, and machine learning approaches, are currently in use. To facilitate experimental realization of unexplored structures their proper characterization is needed. It can be successfully realized in computational simulations that have become an effective tool in the prediction and characterization of unknown compounds (12,13). A combination of density functional theory (DFT) calculations and machine learning algorisms allows one to discover exotic low-dimensional structures. A broad range of computational techniques has been utilized to describe the crystal structure of potentially superhard boron-rich $MoB_x$ phases (14). Later, using the evolutionary algorithm, five new superhard ternary compounds in the W−Mo−B system have been predicted at different temperatures, and the composition−temperature phase diagrams have been plotted (15). A thin film of NaCl on the (110) diamond surface has been crystallized in the experiment based on a theoretical guidance provided by the *ab initio* calculations combined with an evolutionary algorithm (16). Data-driven high-throughput investigations have led to identification of 8 binary and 20 ternary non-vdW potentially synthesizable candidates with the hematite and ilmenite type structures (8). Recently, computationally guided high-throughput synthesis has been used to explore the Zn–V–N phase space, resulting in the synthesis of a novel ternary nitride $Zn_2VN_3$ film (17).

Wide band gap wurtzite $Zn_2VN_3$ thin films exhibit *p*-type conductivity, charge carrier concentration of $\sim 10^{17}$ cm$^{-3}$, and promising Hall mobility of about 80 cm$^2$/V·s (17). Furthermore, charge carrier concentration in $Zn_2VN_3$ is controlled by Zn/V ratio. The favorable set of optical and electronic properties makes $Zn_2VN_3$ thin films an interesting candidate for hole-selective contacts and hole transport layer applications in solar cells. Tunability of carrier concentration in $Zn_2VN_3$ may also be used to fabricate solar cells with back surface field to facilitate charge transport. Demonstration of epitaxial stabilization of sputter-deposited $Zn_2VN_3$ thin films at low synthesis temperatures (<200°C) and chemical stability of the nitride material may be suitable for application in tandem perovskite-Si solar cells, in which a diffusion barrier is desired to protect the bottom cell (18). However, the synthesized thin films are characterized by cation disorder. Potentially this can be avoided in 2D $Zn_2VN_3$, which may open up additional functionalities for this novel semiconductor material.

In this work, following the recent prediction and synthesis of wurtzite $Zn_2VN_3$ thin films, the existence of 2D $Zn_2VN_3$ is investigated. A comprehensive study of its dynamic and thermal stability as well as the grows mechanism is reported. A thorough characterization of 2D $Zn_2VN_3$ is conducted, including identification of its opto-electronic and mechanical properties. In addition, structural defects in 2D $Zn_2VN_3$ and its environmental stability are explored. The reported study both predicts a new 2D material and offers its characterization and possible applications.

A monolayer of 2D $Zn_2VN_3$ is obtained from the recently predicted and synthesized bulk $Zn_2VN_3$ (17) by cutting it along the (001) direction, as shown in Figure 1a. The top and side views of the 2D $Zn_2VN_3$ unit cell are also presented in Figure 1a, and the conventional cell is shown in Figure S1a. The unit cell of 2D $Zn_2VN_3$ consisting of 24 atoms (8 Zn atoms, 12 N atoms and 4 V atoms) stabilizes in a 2D orthorhombic lattice with the space group 36 $Cmc2_1$ and the lattice parameters are $a = 5.72$ and $b = 5.63$ Å (see cif file in SI). The electronic localization function (ELF) reflects the degree of charge localization in the real space, where 0 represents a free electronic state while 1 represents a perfect localization. The calculated ELF for 2D $Zn_2VN_3$ with the isosurface value of 0.2 (Figure 1a) reflects electron density. The electron localization basin is spherical and completely migrates to the Zn atom. All basins surround the respective cores, suggesting an ionic bond in 2D $Zn_2VN_3$. The existence of strong ionic bonds in 2D $Zn_2VN_3$ suggests its high stability against formation of most point defects (19). The calculated phonon dispersion spectra of 2D $Zn_2VN_3$ along the high symmetry path of the Brillouin zone (Figure 1b) shows its kinetic stability, as the transverse, longitudinal and out-of-plane $z$-direction acoustic modes have real frequencies and display normal linear dispersion around the $\Gamma$ point. Thermal stability of 2D $Zn_2VN_3$ is confirmed via AIMD simulations showing that the structure remains stable after 5 ps at 300 K (Figure S1b and Movie 1)

The majority of 2D materials are exfoliated from powders/thin films or designed via special methods such as chemical vapor deposition (20). Feasibility of exfoliation of 2D $Zn_2VN_3$ needs to be evaluated. The calculated binding energy that needs to be overcome to achieve exfoliation of a 2D $Zn_2VN_3$ monolayer from bulk $Zn_2VN_3$ is 2.83 eV, which is ~20 times higher than that of graphene (Figure S2). The exfoliation energy $\Delta E_{exf}$ for 2D $Zn_2VN_3$, that is, the binding energy between layers in the bulk, is computed as the energy difference between the relaxed 2D and bulk systems (21),

$$\Delta E_{\text{exf}} = \frac{E_{2D} - E_{bulk}}{A}. \quad (1)$$

Here, $E_{2D}$ and $E_{bulk}$ correspond to the total energies of the optimized 2D and bulk $Zn_2VN_3$, respectively, and A is the in-plane surface area according to the optimized bulk $Zn_2VN_3$. $\Delta E_{exf}$ of 2D $Zn_2VN_3$ is found to be 105 meV/Å$^2$. This value is ~5 times higher than the $\Delta E_{exf}$ value of graphene (~20 meV/Å$^2$) (21,22), while it is below the 130-200 meV/Å$^2$ limit proposed for "potentially exfoliable" systems (22,24). Therefore, exfoliation of the 2D $Zn_2VN_3$ monolayer should be possible at certain conditions. The recently reported approach for deposition of bulk $Zn_2VN_3$ suggests that it can be formed from evaporation of $Zn_3N_2$ and VN at a temperature of 390-490 K and in the presence of ionized nitrogen, roughly following the reaction $Zn_3N_2 + VN \rightarrow (N^+) Zn_2VN_3 + Zn$ (evaporation) (17). AIMD simulations conducted in this work show that a $Zn_2VN_3$ structure containing an extra (unevaporated) Zn atom can be formed at 180 K within ~1.3 ps (Movie 2). Therefore, by controlling the Zn evaporation rate and the ionized nitrogen rate it can be possible to utilize chemical vapor deposition for 2D $Zn_2VN_3$ synthesis. During the manufacturing, the thickness of sputter-deposited thin films is increased with sputtering power (for constant deposition time). For example, the thickness of the Mo-incorporated $Cu_2ZnSnS_4$ absorber layer increased with Mo co-sputtering power, as shown in the cross-section scanning transmission electron microscope study (25). More recently, the ternary metal-zinc-nitride thin films of about 235 nm thickness were demonstrated experimentally and implemented in a photodetector device

(26). Proper choice of a substrate is an important step for production of 2D $Zn_2VN_3$. For instance, bulk $Zn_2VN_3$ deposited on glass and sapphire substrates appears to be phase-pure, while the epitaxial stabilization on $Al_2O_3$ (0001) increases its crystallinity and texture (25).

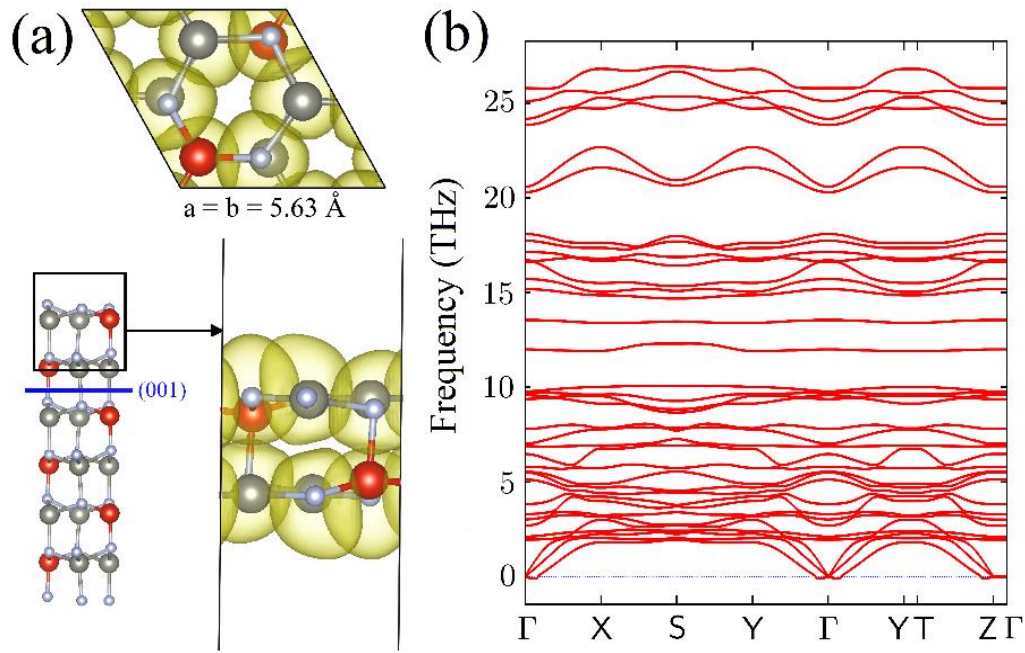

**Figure 1. (a)** Schematic representation of $Zn_2VN_3$ transformation from bulk to 2D, and the 2D $Zn_2VN_3$ unit cell combined with ELF. The Zn, V, and N atoms are colored in gray, red and violet, respectively. **(b)** Phonon dispersion curves of 2D $Zn_2VN_3$.

Figure 2a shows the band structure of 2D $Zn_2VN_3$ obtained using the Heyd–Scuseria–Ernzerhof (HSE06) exchange-correlation functional. For the comparison, the band structure of 2D $Zn_2VN_3$ obtained via the Perdew-Burke-Ernzerhof (PBE) functional under the generalized gradient approximation is plotted in Figure S3a. Based on the HSE06 calculations, it is found that 2D $Zn_2VN_3$ is a semiconductor with an indirect band gap of 2.75 eV and a direct band gap of 2.85 eV. Both gaps are slightly higher than in bulk $Zn_2VN_3$ (17). Similarly to its bulk counterpart, 2D $Zn_2VN_3$ is a *p*-type semiconductor with the Fermi level located slightly above the valence band edge. For the indirect gap, the conduction band minimum (CBM) is located in the vicinity of the Γ point, while the valence band maximum (VBM) is located at the S point. In the case of the direct gap, both CBM and VBM are located at the S point. The partial density of states (PDOS) of 2D $Zn_2VN_3$ (Figure S3b) demonstrates that *d* states of V atoms give the main contribution to the CBM, while the VBM is formed by *p* states of N atoms. From the band structure and PDOS plots the CBM localization is noticed. Such localization of states around the CBM is commonly observed in materials due to cation disorder (27, 28). Localization on the *d* states of V atoms has been found in bulk $Zn_2VN_3$ (17). It is also supposed that the origin of the CBM localization in 2D $Zn_2VN_3$, as reflected in the PDOS plot (Figure S3), is due to the cation disordering.

The optical response of a material at a given frequency can be determined via its frequency-dependent complex dielectric function that can be defined as a function of incident photon energy having real and imaginary parts (29). This function describes the process of light propagation through the material. Light absorption of a material at a given frequency ω can be simply realized by the positive value of the real part of the dielectric function. Therefore, the real part refers to the ability of a material to store the electric energy. The imaginary part of the dielectric function refers to the ohmic resistance of the material. For instance, the imaginary part of the dielectric function of a pure dielectric has a zero value. The real and imaginary parts of the complex dielectric function for 2D $Zn_2VN_3$ are presented in Figure 2b. The real part of the dielectric function for the incident electromagnetic field normal to the 2D $Zn_2VN_3$ surface is positive in the considered region of the electromagnetic spectra from 0 to 16 eV. This suggests that photons propagate through the monolayer. The static dielectric function, which represents the dielectric response of the material to a static electric field, is found to be 1.28. The real dielectric constant has a considerable peak at 4.9 eV with the maximum value of 1.55, and it exhibits a few peaks at higher energies. The imaginary part of the dielectric function shows that 2D $Zn_2VN_3$ absorbs light in the visible and ultraviolet regions. Considering the electric and optical properties of the predicted 2D $Zn_2VN_3$, it can be regarded as a transparent material for the visible lights and a useful shielding material in ultraviolet region.

Figure 2c shows the diagram of the 2D $Zn_2VN_3$ work function in comparison with other 2D materials and bulk metals possessing the highest known work function values. The work function of 2D $Zn_2VN_3$ is high, 5.27 eV, comparable to that of borophene (30). The high work function of 2D $Zn_2VN_3$ can be attributed to the nature of its atomic states around the Fermi level consisting of not only the out-of-plane $p_z$ states but also the in-plane $p$ hybridized states (Figure S4). Thus, the ionization of 2D $Zn_2VN_3$ requires significant energy and is comparable to that of, for example, borophene.

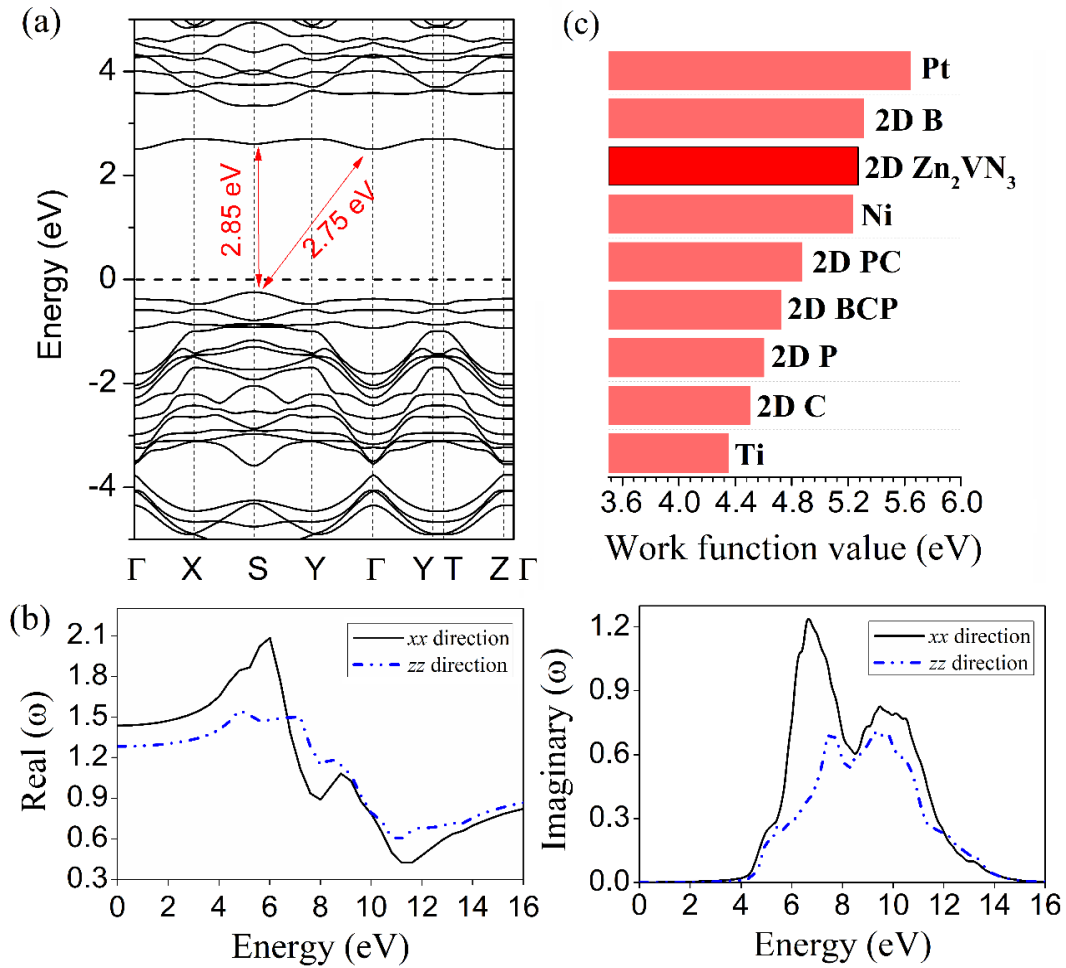

**Figure 2. (a)** Band structure of 2D $Zn_2VN_3$ calculated by the HSE approach. **(b)** Real and imaginary parts of dielectric function versus energy for 2D $Zn_2VN_3$. **(c)** Comparison of the work function of 2D $Zn_2VN_3$ with those of common 2D materials and bulk metals.

Mazdziarz formulated the mechanical stability of 2D materials acquired in rectangular lattices as follows (31):

$$\frac{1}{2}(C_{11}+ C_{22} + \sqrt{4C_{12}^2 - (C_{11} - C_{22})^2}\,) > 0$$

$$\frac{1}{2}(C_{11}+ C_{22} - \sqrt{4C_{12}^2 - (C_{11} - C_{22})^2}\,) > 0 \quad (2)$$

$$C_{66} > 0$$

The above presented criteria are met in the case of 2D $Zn_2VN_3$ confirming its mechanical stability. The calculated elastic constants $C_{ij}$ for 2D $Zn_2VN_3$ are collected in Table S1.

Mechanical properties of 2D $Zn_2VN_3$ such as Young's modulus, Poisson's ratio and shear modulus are also considered. Young's modulus of 2D $Zn_2VN_3$ in the x and y directions is calculated as (32,33):

$$E_{[x]}=\frac{C_{11}C_{22}-C_{12}^2}{C_{11}}, \text{ and } E_{[y]}=\frac{C_{11}C_{22}-C_{12}^2}{C_{22}} \quad (3)$$

Shear modulus of 2D $Zn_2VN_3$ is calculated as (30,31):
$$G = C_{66} \quad (4)$$

Poisson's ratio of 2D $Zn_2VN_3$ in the x and y directions is calculated as (30,31):

$$v_{[x]}=\frac{C_{12}}{C_{11}}, \text{ and } v_{[y]}=\frac{C_{12}}{C_{22}} \quad (5)$$

Figure 3 presents the spatial dependence of Young's modulus, shear modulus and Poisson's ratio for 2D $Zn_2VN_3$. These parameters are almost direction independent. The values of Young's modulus of 2D $Zn_2VN_3$ in the x and y directions are found to be ~99.7 N/m. Shear modulus of 2D $Zn_2VN_3$ is found to be 37.6 N/m. Poisson's ratio of 2D $Zn_2VN_3$ in both x and y directions is found to be 0.375. For comparison, 2D $Zn_2VN_3$ has lower stiffness and higher elasticity relative to graphene (34).

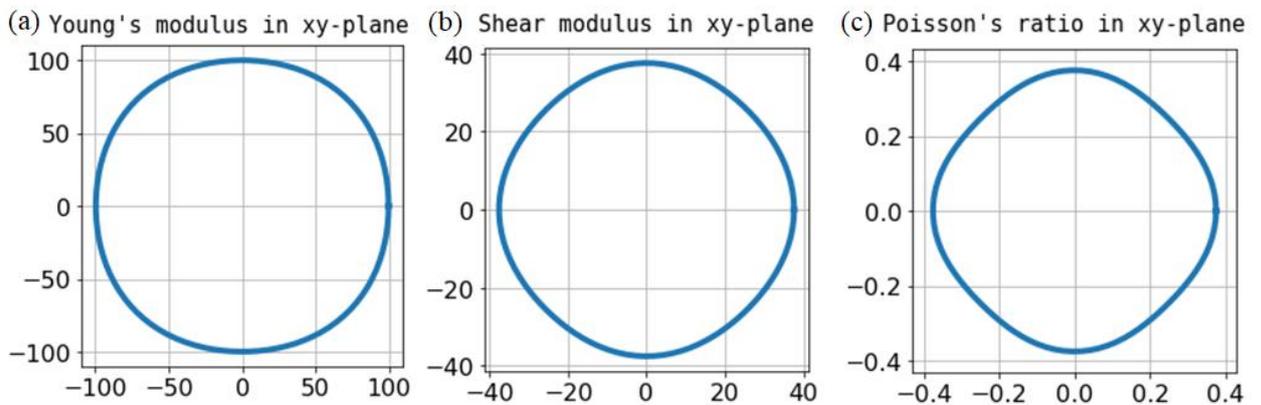

**Figure 3.** Spatial dependencies of **(a)** Young's modulus (N/m), **(b)** shear modulus (N/m), and **(c)** Poisson's ratio for 2D $Zn_2VN_3$.

2D $Zn_2VN_3$ can be switched from a direct band gap semiconductor to an indirect band gap semiconductor and its band gap size can be increased/decreased up to ~50% via strain engineering. Although the PBE functional underestimates the band gap compared to the HSE functional, PBE (Figure S3a) and HSE (Figure 2a) display similar trends in the band alignment in the band structure plots. Thus, the effect of strain engineering on the band structure of 2D $Zn_2VN_3$ can be studied using the PBE approach. Figure 4a shows changes in the band gap of 2D $Zn_2VN_3$ under strain applied along the surface plane. The applied strain is in the range from -10% (compressive strain) to 10% (tensile strain) that can be realized experimentally (35-40). The band gap of 2D $Zn_2VN_3$ decreases rapidly from 1.57 eV to 0.76 eV when the compressive strain increases from 0 to 10%. As the VBM shifts from the S point to the $\Gamma$ point (Figure S5), an indirect-direct band gap transition is observed in 2D $Zn_2VN_3$ under the compressive strain higher than 6%. A rapid decrease of the 2D $Zn_2VN_3$ band gap from 1.57 eV to 1.07 eV is observed with the increase of the tensile strain from 0% to 8%. The applied tensile strain in the range from 8% to 10% leads to a slight increase of the band gap from 1.07 eV to 1.19 eV (indirect band gap) and 1.22 eV (direct band gap) (Figure S5).

As 2D materials always contain surface defects that may affect their structure stability and performance (41-45), it is necessary to investigate point defects in 2D $Zn_2VN_3$. Several common point defects are found to be stable in 2D $Zn_2VN_3$: a single vacancy of a N atom ($SV_N$); a single vacancy of a Zn atom ($SV_{Zn}$); a single vacancy of a V atom ($SV_V$); in-plane and out-of-plane double vacancies of one V atom and one N atom ($DV_{V-N}$); and in-plane and out-of-plane double vacancies of one Zn atom and one N atom ($DV_{Zn-N}$). SV defects can be created by removing one Zn, V or N atom from the 2D $Zn_2VN_3$ surface. The $DV_{V-N}$ and $DV_{Zn-N}$ defects in the 2D $Zn_2VN_3$ surface are formed when V and N or Zn and N atoms are removed from the surface (in-plane DV) or from the plane perpendicular to the surface (out-of-plane DV) of 2D $Zn_2VN_3$. The stability of point defects in 2D $Zn_2VN_3$ is evaluated based on their formation energy, $E_{form}$. $E_{form}$ of the considered stable defects in 2D $Zn_2VN_3$ are collected in Table 1. According to Table 1, the $SV_{Zn}$ and $SV_N$ defects have the lowest $E_{form}$ of 4.27 eV and 5.27 eV, respectively. The out-of-plane $DV_{Zn-N}$, in-plane $DV_{Zn-N}$, $SV_V$, in-plane $DV_{V-N}$, and out-of-plane $DV_{V-N}$ defects have comparably high formation energies of 7.83 eV, 8.54 eV, 10.25 eV, 10.96 eV, and 11.92 eV, respectively. $E_{form}$ of SVs in 2D $Zn_2VN_3$ is comparable to that in graphene (~7.50 eV) (46) and $MoS_2$ (2.10-6.20 eV) (47), while the formation of the DVs in 2D $Zn_2VN_3$ is less favorable than in graphene (~8.0 eV) (46) and $MoS_2$ (~4.0 eV) (47).

**Table 1.** $E_{form}$ (eV) of point defects in 2D $Zn_2VN_3$.

| $SV_N$ | $SV_{Zn}$ | $SV_V$ | in-plane $DV_{V-N}$ | out-of-plane $DV_{V-N}$ | in-plane $DV_{Zn-N}$ | out-of-plane $DV_{Zn-N}$ |
|---|---|---|---|---|---|---|
| 5.27 | 4.27 | 10.25 | 10.96 | 11.92 | 8.54 | 7.83 |

Proper identification and classification of defects is a key capability of atomically resolved scanning tunneling microscopy (STM) (48,49). To facilitate experimental needs it is possible to utilize DFT simulated STM images for the differentiation of point defects in 2D $Zn_2VN_3$. The atomic structures and corresponding STM images of the pristine and defect-containing 2D $Zn_2VN_3$ are presented in Figures S6a-d and S7a-d. The STM images of 2D $Zn_2VN_3$ containing the considered defects correlate with the corresponding atomic structures, and the defects are easily identified. The STM image in Figure S6b (right panel) shows that the $SV_N$ defect in 2D $Zn_2VN_3$ can be identified by the triangle formed with two bright spots and one dark spot, arising from two Zn atoms and one V atom, with one N atom missing inside the triangle. Similarly, according to

Figures S6c and d (right panels), the $SV_V$ and $SV_{Zn}$ defects in 2D $Zn_2VN_3$ appear as triangles formed of three small dark spots characterizing three N atoms, with V or Zn atoms missing inside the triangles. The in-plane $DV_{V-N}$ defect in 2D $Zn_2VN_3$ is presented in Figure S7a. Here, two semi-hexagons, formed of two bright spots (Zn atoms) and two dark spots (N atoms) each, have two missing atoms (one V atom and one N atom) at their border. The out-of-plane $DV_{V-N}$ defect in 2D $Zn_2VN_3$ (Figure S7b) is characterized by missing V and Zn atoms in-between five big bright spots (Zn atoms) and one dark spots (N atom) slightly shifted from their positions compared to the perfect case. The in-plane $DV_{Zn-N}$ defect in 2D $Zn_2VN_3$ is seen in Figure S7c as two missing atoms (Zn and N) inside a square formed of three dark spots (two N atoms and one V atom). The out-of-plane $DV_{Zn-N}$ defect in 2D $Zn_2VN_3$ (Figure S7d) may be identified as missing atoms within a triangle formed of two bright spots and one dark spot due to two Zn atoms and one V atoms. This pattern is similar to the STM image the $SV_N$ defect. To deeper understand the changes in the electronic structure of 2D $Zn_2VN_3$ induced by the defects, the density of states resolved in space, known as local density of states (LDOS), calculated for peripheral atoms in the defect core and for atoms far from the defect core, are shown in Figures S8 and S9. It is found that the defects induce significant changes in the electronic structure of 2D $Zn_2VN_3$. Such changes facilitate defect identification via photoemission spectroscopy techniques.

Figure 4b presents the temperature-depended surface density of point defects in 2D $Zn_2VN_3$. The $SV_N$ and $SV_{Zn}$ defects possess significantly higher surface densities compared to the other defects found to be stable in 2D $Zn_2VN_3$. The surface density of the $SV_V$, out-of-plane $DV_{Zn-N}$, in-plane $DV_{Zn-N}$, in-plane $DV_{V-N}$, and out-of-plane $DV_{V-N}$ defects in 2D $Zn_2VN_3$ is slightly lower than that in graphene (50) and $MoS_2$ (51), making their formation less energetically favorable. On the other hand, the surface density of the $SV_N$ and $SV_{Zn}$ defects in 2D $Zn_2VN_3$ is comparable to that of the SV defects in graphene (50) and $MoS_2$ (51).

Structural degradation of 2D materials can also be caused by their interaction with the humid environment, particularly, with $H_2O$ and $O_2$ molecules contained in the air. Therefore, the interaction of the $H_2O$ and $O_2$ molecules with the 2D $Zn_2VN_3$ surface is further evaluated. As it has been shown previously, for most of 2D materials, oxidation is the most dangerous process that can lead to degradation of their surface (52-54), while $H_2O$-saturated surfaces can exhibit higher stability as such saturation prevents the oxidation (55,56). It is found that the adsorption energy, $E_{ads}$, of $O_2$ on the 2D $Zn_2VN_3$ surface is as high as -0.14 eV, while $E_{ads}$ of $H_2O$ (-0.49 eV) is ~3.5 lower (Figure S10). Hence, the adsorption of $H_2O$ on 2D $Zn_2VN_3$ is more favorable compared to $O_2$. It should be noted, that according to the LDOS plots (Figure S11) remarkable changes in $H_2O$ and $O_2$ molecular states upon their interaction with the 2D $Zn_2VN_3$ surface are observed. Therefore, the kinetic analysis of $H_2O$ and $O_2$ splitting on the 2D $Zn_2VN_3$ surface is further conducted. Figure 4c presents the result from the CI-NEB calculation. The energy barrier, $E_b$, for the $H_2O$ and $O_2$ molecule dissociation on 2D $Zn_2VN_3$ is found to be 2.82 eV and 2.04 eV, respectively. Considering the obtained high values of $E_b$ for the dissociation of $H_2O$ and $O_2$ on the 2D $Zn_2VN_3$ surface, which are comparable to those of the $H_2O$ and $O_2$ molecule dissociation on InSe (57), and based of the $E_{ads}$ analysis, it can be concluded that 2D $Zn_2VN_3$ possesses high structural integrity under environmental conditions. A discussion on the potential application of 2D $Zn_2VN_3$ to water splitting is presented in Supporting Information, and the calculated VBM and CBM positions of 2D $Zn_2VN_3$ with the redox potential of $H_2O$ and oxidation levels are shown in Figure S12.

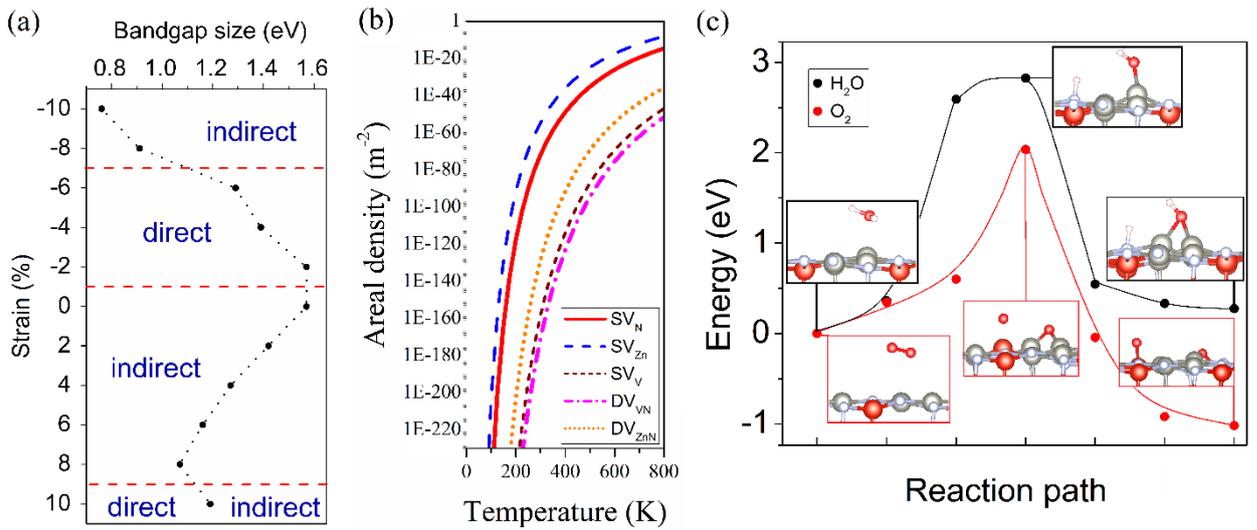

**Figure 4. (a)** Band gap of 2D $Zn_2VN_3$ as a function of strain. **(b)** Surface density of point defects in 2D $Zn_2VN_3$ as a function of temperature. **(c)** Activation barriers for $H_2O$ and $O_2$ molecule splitting on 2D $Zn_2VN_3$.

In summary, a new material, a 2D analog of the recently predicted and synthesized ternary nitride semiconductor $Zn_2VN_3$ (17), is predicted computationally. The fabrication of 2D $Zn_2VN_3$ is highly possible due to is relatively low exfoliation energy of 105 meV/Å$^2$. It is also proposed that the chemical vapor deposition approach can be utilized for the synthesis of 2D $Zn_2VN_3$ similarly to the bulk $Zn_2VN_3$ (17) and $Cu_2ZnSnS_4$ thin film (25) manufacturing. The environmental stability of 2D $Zn_2VN_3$ should be high due to resistivity of its surface to oxygen and defect formation. 2D $Zn_2VN_3$ has an indirect band gap of 2.75 eV, which can be tuned up to 50% under applied stains. 2D $Zn_2VN_3$ possesses a high work function of 5.27 eV, absorbs visible and ultraviolet light, and exhibits moderate mechanical properties. All this makes 2D $Zn_2VN_3$ a good candidate for application in opto-electronic and straintronic devices.

**Computational Methods**

The spin-polarized first-principles calculations were performed within the framework of density functional theory as implemented in the plane-wave the Vienna Ab initio Simulation Package (VASP) (58). The Perdew-Burke-Ernzerhof functional (PBE) (59) under the generalized gradient approximation and the HSE06 hybrid exchange-correlation functional (60) were used. Dispersive interactions were included using the van der Waals corrected functional (61). The geometry optimization was stopped once the atomic forces and total energy values were smaller than $10^{-4}$ eV/Å and $10^{-8}$ eV, respectively. The plane-wave cut-off energy was set to 520 eV. The periodic boundary conditions were applied for the two in-plane transverse directions. A vacuum space of 25 Å was introduced to the direction perpendicular to the surface plane.

The electron localization function (ELF) was calculated to obtain the distribution of electrons in 2D $Zn_2VN_3$. The Phonopy code associated with VASP was used for the simulation of the phonon spectrum (62). The 3×3×1 supercell of 2D $Zn_2VN_3$ was used to perform the calculations based on finite displacement methods with the atomic displacement distance of 0.01 Å. The dielectric function of 2D $Zn_2VN_3$ was calculated based on the TD-HSE06 approach (63). Ab initio molecular dynamics simulations controlled by the Nose–Hoover thermostat were performed for 5 ps at the temperature of 300 K and with a time step of 1.0 fs (64).

The stress-strain relation (65) was used to calculate the components of the stiffness matrix from which the Young's modulus, shear modulus, and Poisson's ratio were obtained and directional dependencies of these quantities were defined using the ELATE software for analysis of elastic tensors (66).

To consider point defects in 2D $Zn_2VN_3$ a supercell composed of 3×3×1 unit cells (36 Zn, 18 V and 54 N atoms) was created to avoid non-physical interactions between periodic images while keeping affordable computational demand. Under such conditions, the concentration of MV defects was 0.93% and the concentration of DV defects was 1.85%. The Tersoff-Hamann approach was implemented to simulate Scanning Tunneling Microscope (STM) images of pure and defect-containing 2D $Zn_2VN_3$ (67).

The formation energy $E_{form}$ of point defects in 2D $Zn_2VN_3$ was calculated as

$$E_{form} = E_{defect} - E_{pure} + N_{Zn} \cdot E_{Zn} + N_V \cdot E_V + N_N \cdot E_N, \tag{6}$$

where $E_{defect}$ and $E_{pure}$ are the total energies of pure and defect-containing 2D $Zn_2VN_3$, $E_{Zn}$, $E_V$ and $E_N$ are the energies of single Zn, V and N atoms, and $N_{Zn}$, $N_V$ and $N_N$ correspond to the number of the removed Zn, V and N atoms.

The surface density of defects $N_d$ in 2D $Zn_2VN_3$, at a finite temperature, was calculated according to the Arrhenius equation:

$$N_d = N_{pure}\, e^{-E_{form}/(k_B T)}, \tag{7}$$

where $N_{pure}$ is the surface density of atoms in pure 2D $Zn_2VN_3$, $k_B$ is the Boltzmann constant, and $T$ is temperature. Note that only defects presented at the surface were shown.

The climbing image–nudged elastic band (CI-NEB) method was used to obtain the reaction pathway of the $H_2O$ and $O_2$ molecules on the 2D $Zn_2VN_3$ surface (68).

**ASSOCIATED CONTENT**

**Notes**

The authors declare no competing financial interest.

**ACKNOWLEDGMENTS**

S.A.Sh. is thankful for the funding provided by the Russian Science Foundation (grant No. 21-71-10129). E.A.K. acknowledges the support of Grant NSh-4320.2022.1.2 of the President of the Russian Federation for state support of young Russian scientists - candidates of sciences and doctors of sciences. A.A.K. is grateful for financial support to the Ministry of Science and Higher Education of the Russian Federation within the framework of the state task of the USATU (No. 075-03-2022-318/1) of the youth research laboratory «Metals and Alloys under Extreme Impacts». Authors acknowledge Peter the Great Saint Petersburg Polytechnic University Supercomputer Center "Polytechnic" and Joint Supercomputer Center of the Russian Academy of Sciences for computational resources. A.A.K. grateful Dr. Siarhei Zhuk for useful discussions on the synthesis and application of Zn–V–N compounds. O.V.P. acknowledges support of the US National Science Foundation (grant CHE-1900510).

Supporting Information

**Section 1. Conventional cell of 2D Zn₂VN₃**

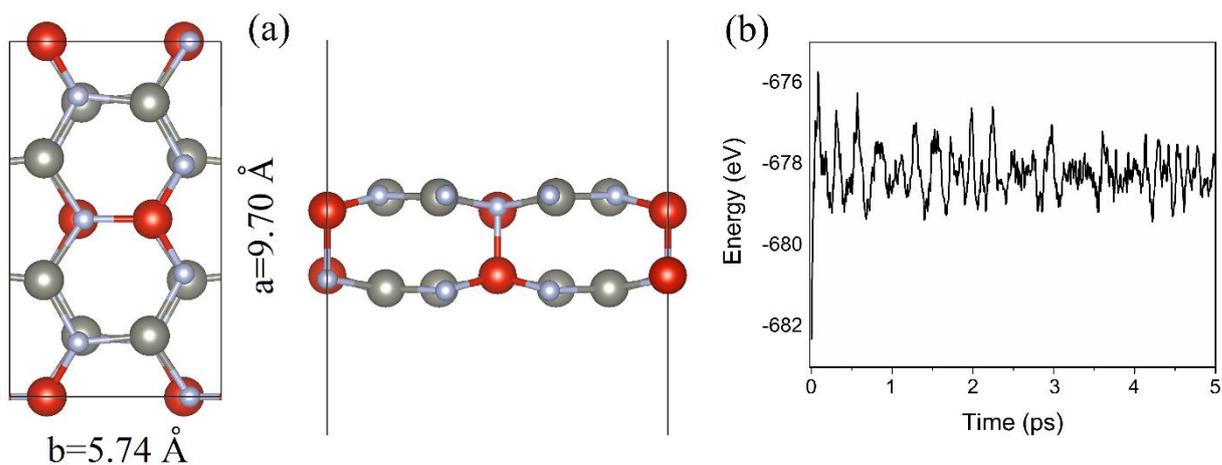

**Figure S1.** (**a**) Conventional cell of 2D Zn₂VN₃. (**b**) Energy fluctuation of the 2D Zn₂VN₃ system from AIMD calculations at 300 K.

**Section 2. Binding energy of 2D Zn₂VN₃**

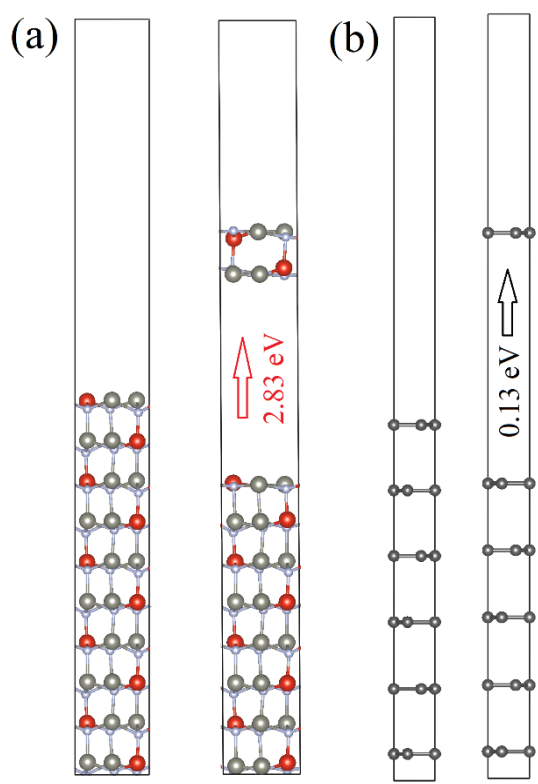

**Figure S2.** Binding energy required for exfoliation of (**a**) 2D Zn₂VN₃ and (**b**) graphene.

**Section 3. GGA band structure and PDOS of 2D Zn$_2$VN$_3$**

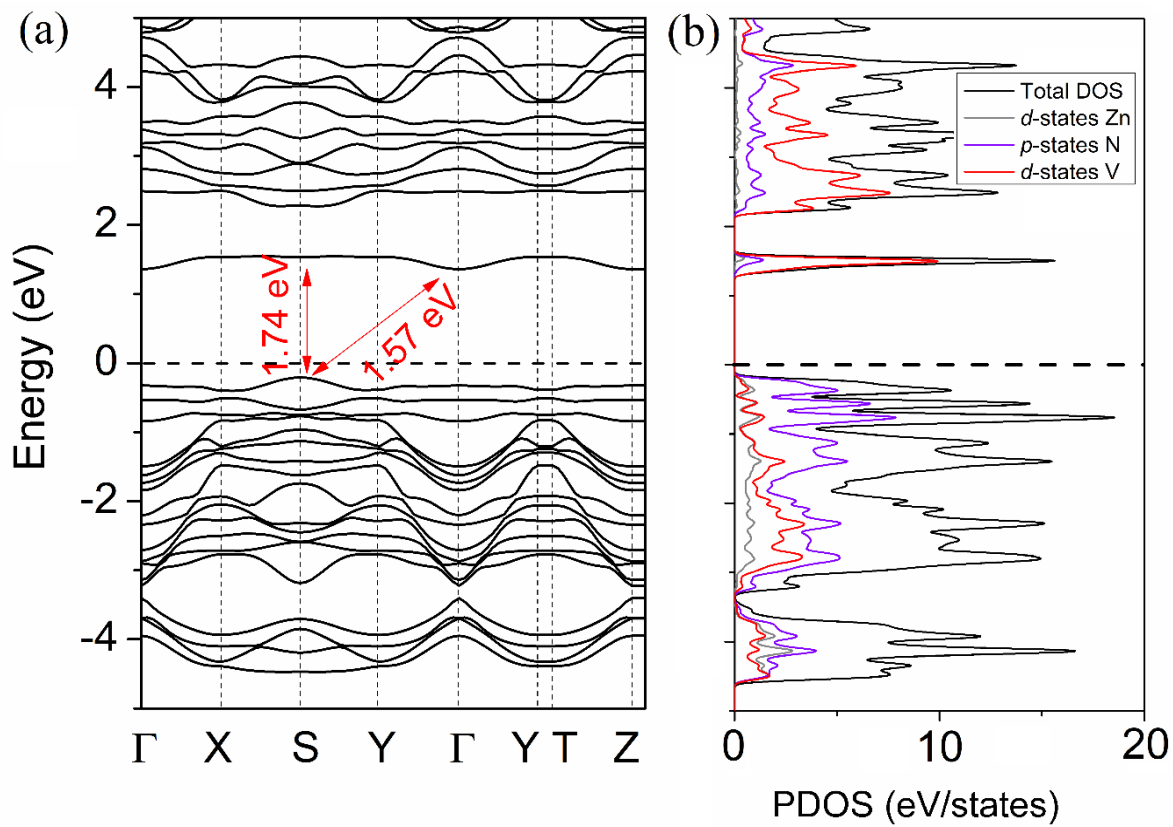

**Figure S3. (a)** Band structure and **(b)** PDOS of 2D Zn$_2$VN$_3$ obtained via the Perdew-Burke-Ernzerhof (PBE) exchange-correlation functional.

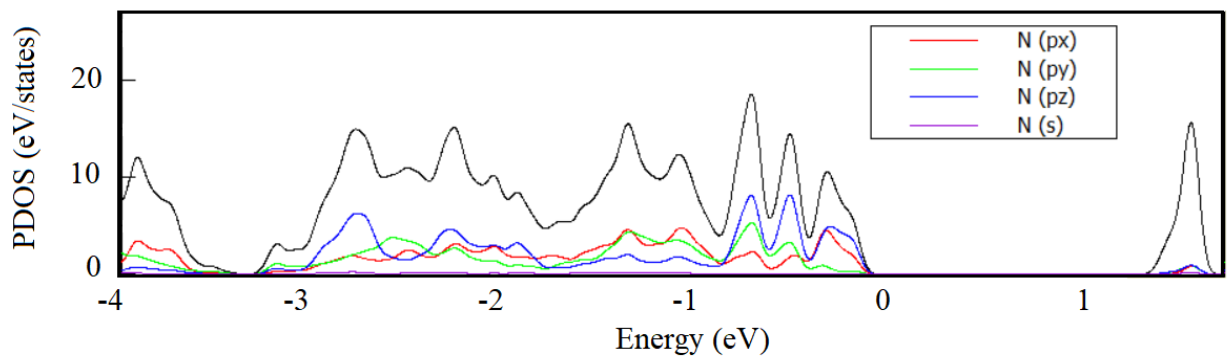

**Figure S4.** PDOS of 2D Zn$_2$VN$_3$.

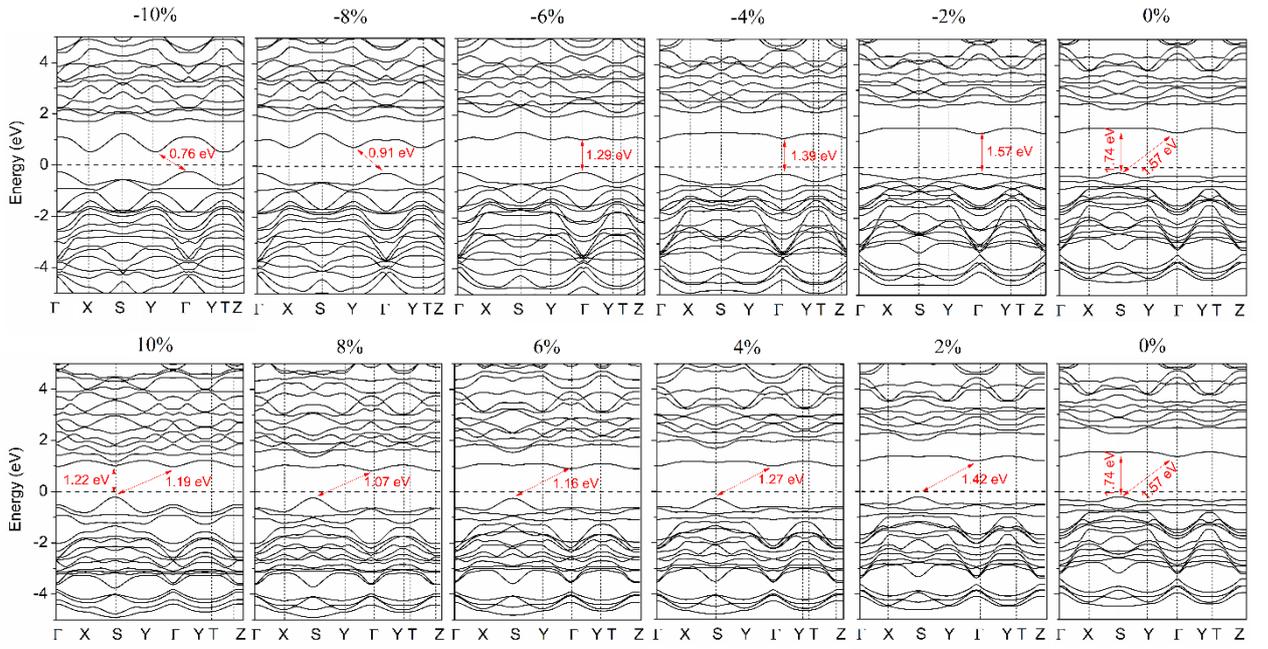

**Figure S5.** Band structure of 2D $Zn_2VN_3$ under compressive (upper row) and tensile (lower row) strain obtained via the PBE approach.

**Section 4. The calculated elastic constants $C_{ij}$ for 2D $Zn_2VN_3$**

**Table S1.** The calculated elastic constants $C_{ij}$ for 2D $Zn_2VN_3$.

| | |
|---|---|
| $C_{11}$, N/m | 117 |
| $C_{22}$, N/m | 117 |
| $C_{12}$, N/m | 44.5 |
| $C_{44}$, N/m | 37.6 |

**Section 5. Atomic stricture, STM images and LDOSs of defect-containing 2D Zn$_2$VN$_3$**

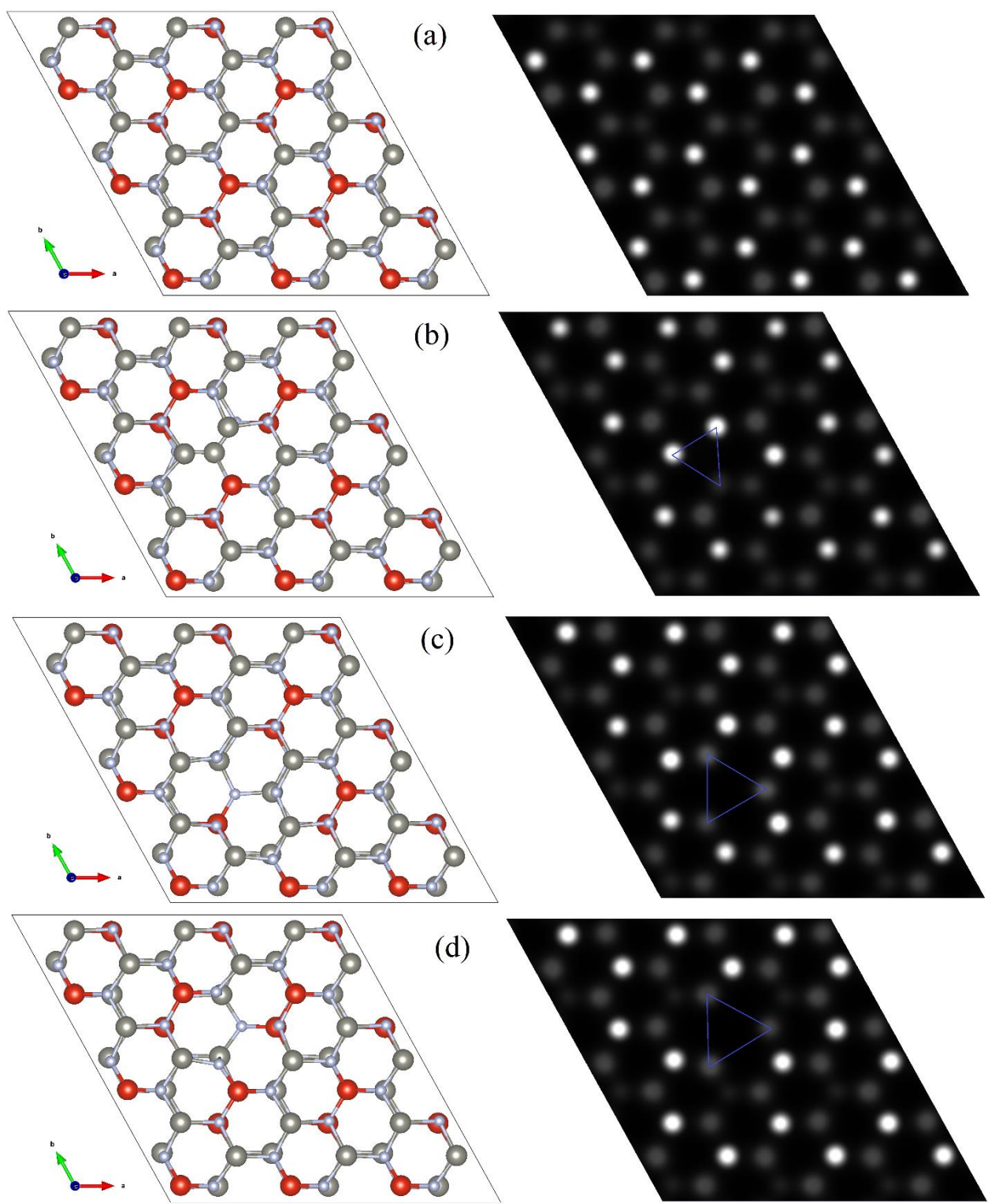

**Figure S6.** Atomic stricture (left) and STM image (right) of **(a)** pure, **(b)** SV$_N$, **(c)** SV$_V$, and **(d)** SV$_{Zn}$ defect-containing 2D Zn$_2$VN$_3$.

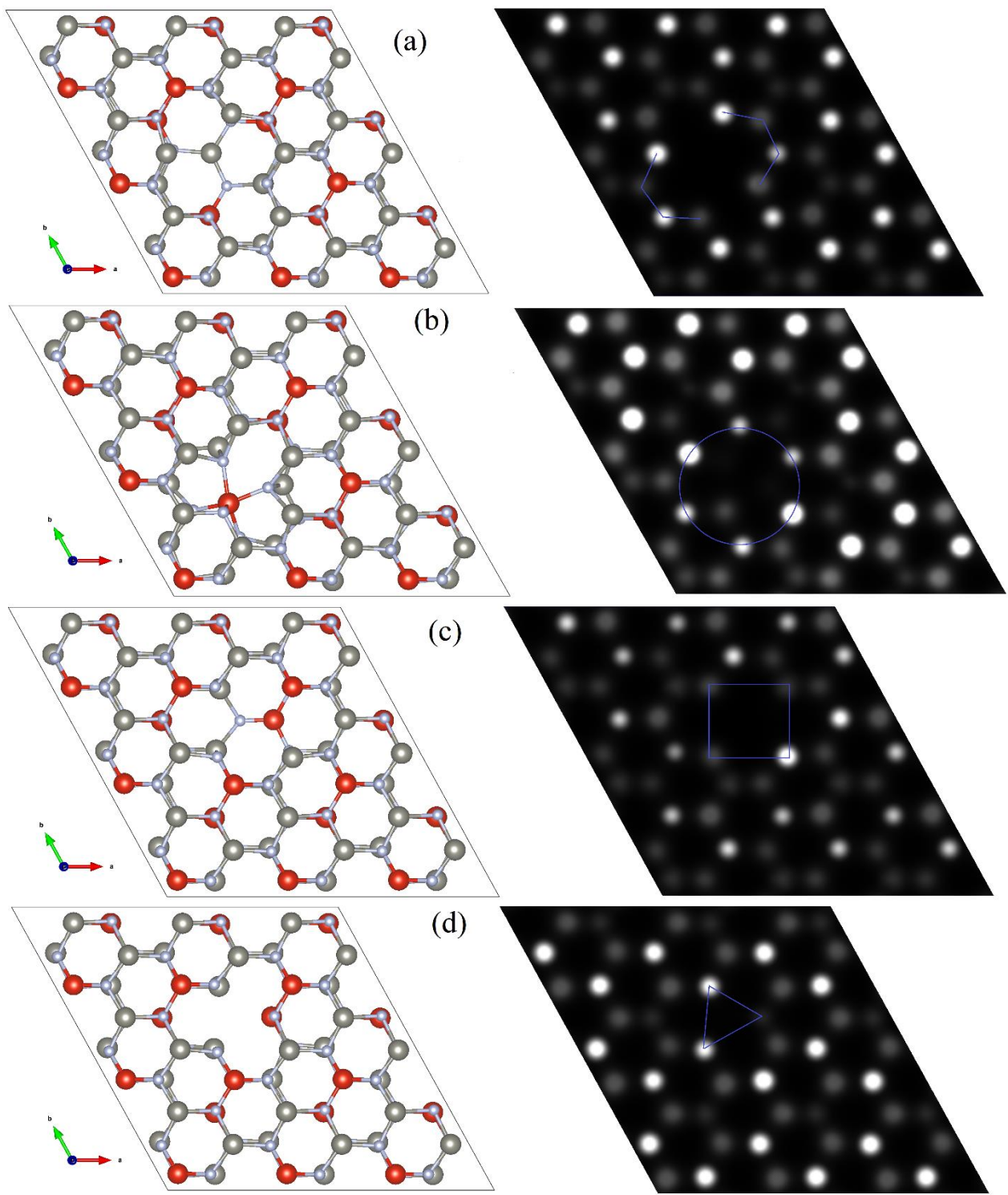

**Figure S7.** Atomic stricture (left) and STM image (right) of **(a)** in-plane $DV_{V\_N}$, **(b)** out-of-plane $DV_{V\_N}$, **(c)** in-plane $DV_{Zn\_N}$, and **(d)** out-of-plane $DV_{Zn\_N}$ defect-containing 2D $Zn_2VN_3$.

To deeper understand the changes in the electronic structure of 2D $Zn_2VN_3$ induced by the defects, the density of states resolved in space, known as local density of states (LDOS), is calculated for peripheral atoms in the defect core and for atoms far from the defect core, as shown in Figures S8 and S9. The states introduced by the $SV_N$ defect are mainly contributed by the V atom (Figure S8a). The changes in the CBM and the VBM depicted in the LDOS plot for the defect (Figure S8b) arise mainly from the $N_1$ and $N_2$ atoms and $N_3$, respectively. The defect-induced states in the vicinity of the VBM in the LDOS plot of the $SV_{Zn}$ system (Figure S8c) mainly originate from the $N_1$, $N_2$ and $N_3$ atoms. In the case of the in-plane $DV_{V\_N}$ defect in 2D $Zn_2VN_3$ (Figure S9a), mainly the $N_1$ and $N_2$ atoms surrounding the defect have partially occupied/unoccupied states contributing into the valence/conduction bands of 2D $Zn_2VN_3$. The $N_1$ atom (Figure S9b) is responsible for the in-gap states appearing in 2D $Zn_2VN_3$ containing the out-of-plane $DV_{V\_N}$ defect. In the case of the in-plane $DV_{Zn\_N}$ in 2D $Zn_2VN_3$, in-gap states appear mainly due to the V atom located in the vicinity of the defect core (Figure S9c). In-gap states in 2D $Zn_2VN_3$, appearing due to the out-of-plane $DV_{Zn\_N}$, are formed by the V, $Zn_2$, and $N_2$ atoms, as shown in Figure S9d.

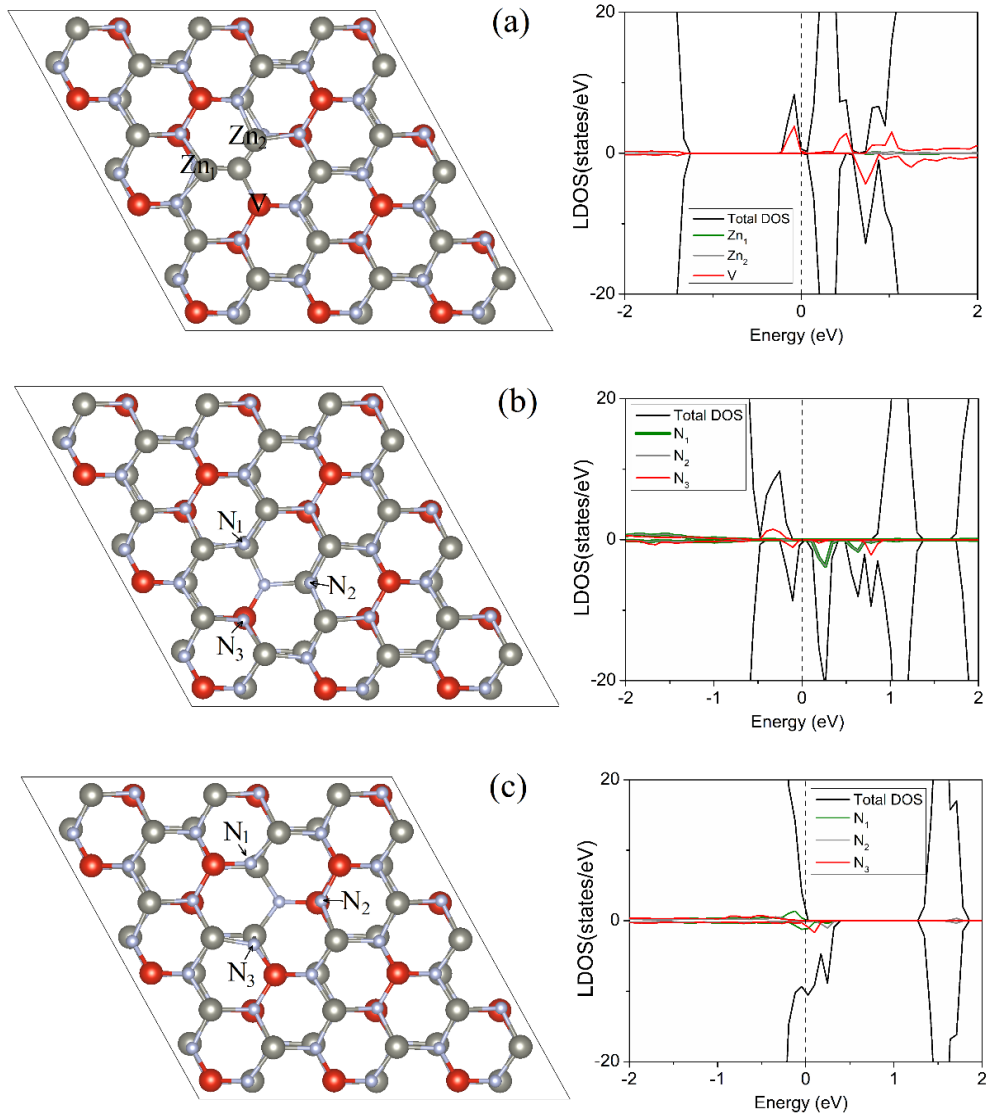

**Figure S8.** Atomic stricture (left) and LDOS (right) of **(a)** $SV_N$, **(b)** $SV_V$, and **(c)** $SV_{Zn}$ defect-containing 2D $Zn_2VN_3$.

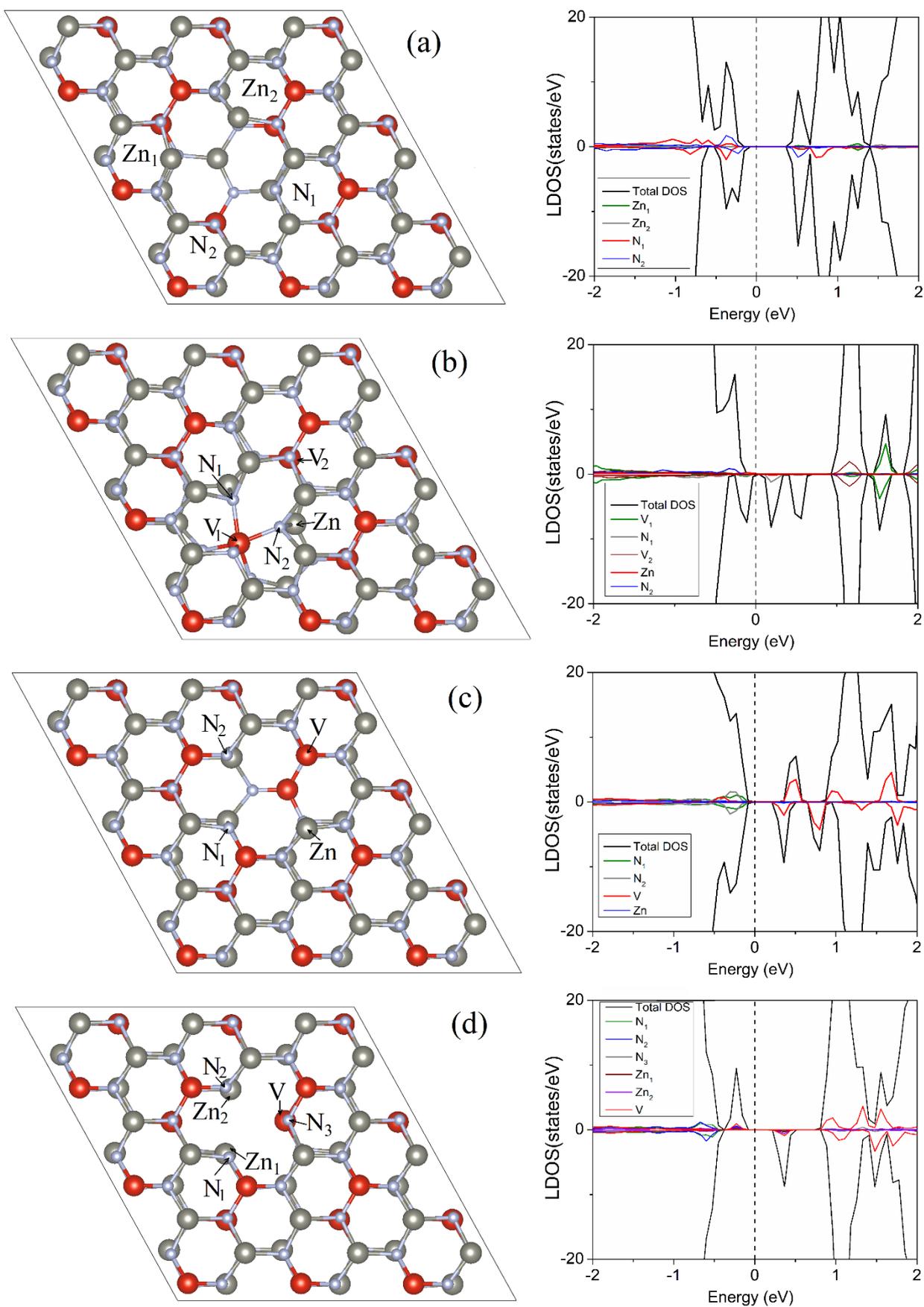

**Figure S9.** Atomic stricture (left) and LDOS (right) of **(a)** in-plane $DV_{V\_N}$, **(b)** out-of-plane $DV_{V\_N}$, **(c)** in-plane $DV_{Zn\_N}$, and **(d)** out-of-plane $DV_{Zn\_N}$ defect-containing 2D $Zn_2VN_3$.

## Section 6. H₂O and O₂ on 2D Zn₂VN₃

According to Figure S10a, the most favorable adsorption position of H$_2$O on 2D Zn$_2$VN$_3$ corresponds to the position of the molecule above the Zn-V bond on the side of the hexagon at the distance of 2.01 Å above the surface of 2D Zn$_2$VN$_3$, and the lowest E$_{ads}$ of H$_2$O on 2D Zn$_2$VN$_3$ is -0.49 eV. Figure S10b shows the most favorable adsorption position of O$_2$ on 2D Zn$_2$VN$_3$. In that case, O$_2$ is in the ring of the hexagon at the distance of 2.73 Å, and the lowest E$_{ads}$ of O$_2$ on 2D Zn$_2$VN$_3$ is -0.14 eV.

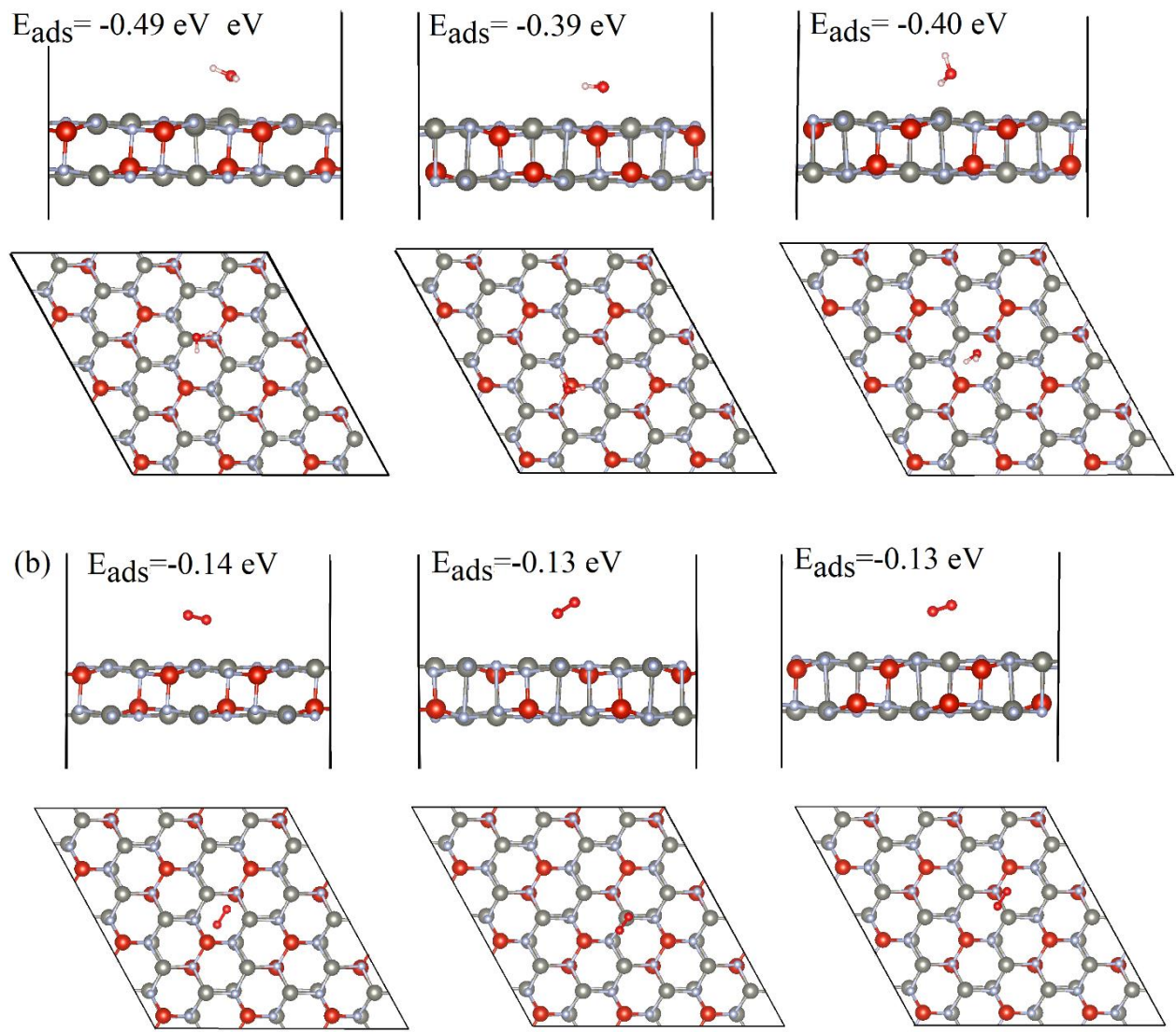

**Figure S10.** Atomic configurations of **(a)** H$_2$O and **(b)** O$_2$ on 2D Zn$_2$VN$_3$.

The three highest occupied molecular orbitals (HOMO) of the $H_2O$ molecule, named according to the irreducible representation of the point group of $H_2O$, are $1b_1$ (HOMO), $3a_1$ (HOMO-1), and $1b_2$ (HOMO-2). According to the LDOS plot (Figure S10a) the $3a_1$ orbital is most broadened due to its favored orbital mixing with the N atoms, confirming the interaction ability of 2D $Zn_2VN_3$ with $H_2O$. The LDOS plot for the $O_2$ molecule on 2D $Zn_2VN_3$ (Figure S10b) reflects additional $O_2$-induced states within the band gap. The half-filled $2\pi$ HOMO state aligns within the valence band maximum and allows the electrons to be excited to the $O_2$ molecule, thereby creating holes in 2D $Zn_2VN_3$. The $2\pi^*$ LUMO state is located above the Fermi level at ~0.90 eV. The presence of the $O_2$-induced states within the band gap of 2D $Zn_2VN_3$ and the non-trivial adsorption/oxidation ability of $O_2$ to 2D $Zn_2VN_3$ can alter its optical and electronic properties.

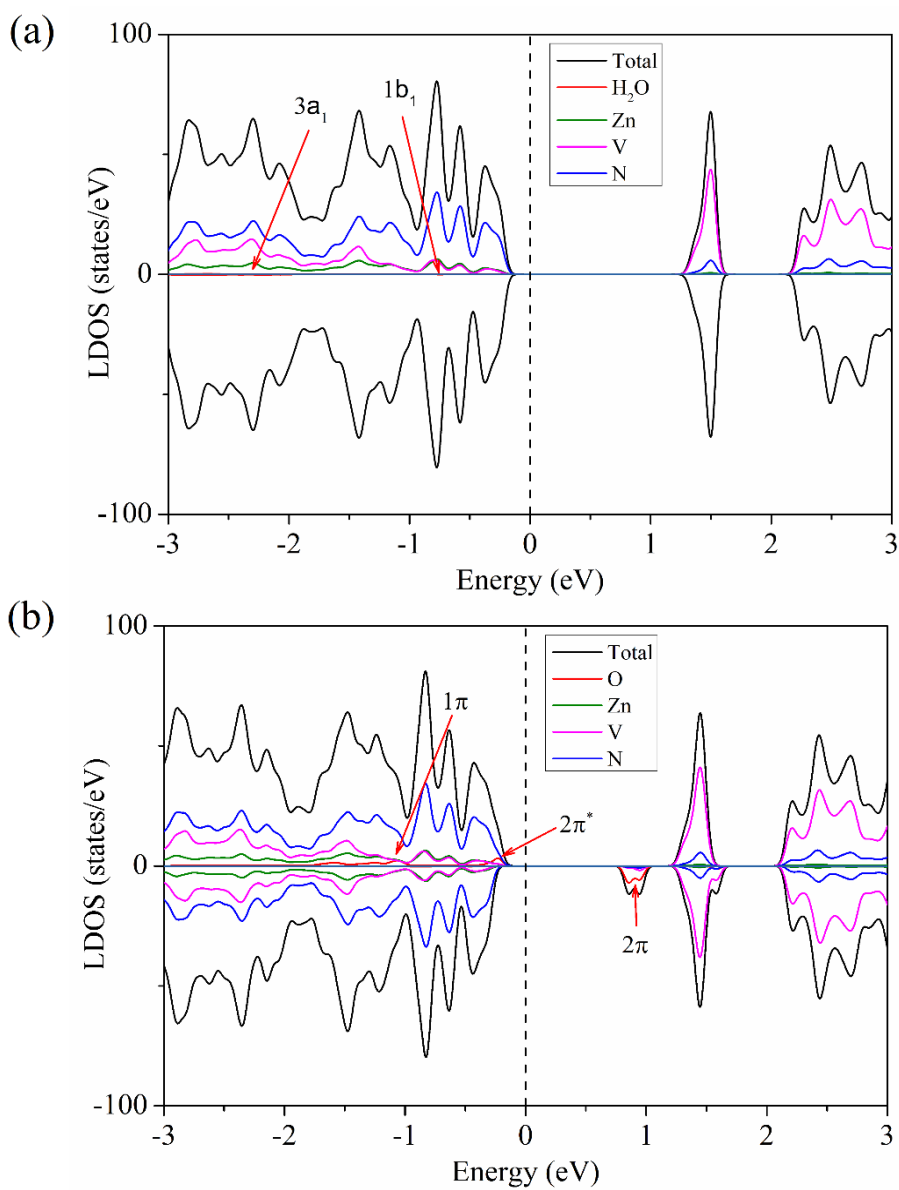

**Figure S11.** LDOS of **(a)** $H_2O$ and **(b)** $O_2$ on 2D $Zn_2VN_3$.

**Section 7.** Water splitting application of 2D $Zn_2VN_3$

Figure S12 shows the calculated VBM and CBM positions of 2D $Zn_2VN_3$ with the redox potential of $H_2O$ and oxidation levels. It is found that the VBM of 2D $Zn_2VN_3$ is below the oxidation potential of $O_2/H_2O$, while the CBM is also lower than the reduction potential of $H_2/H_2O$, which indicates that 2D $Zn_2VN_3$ is suitable only for oxygen production.

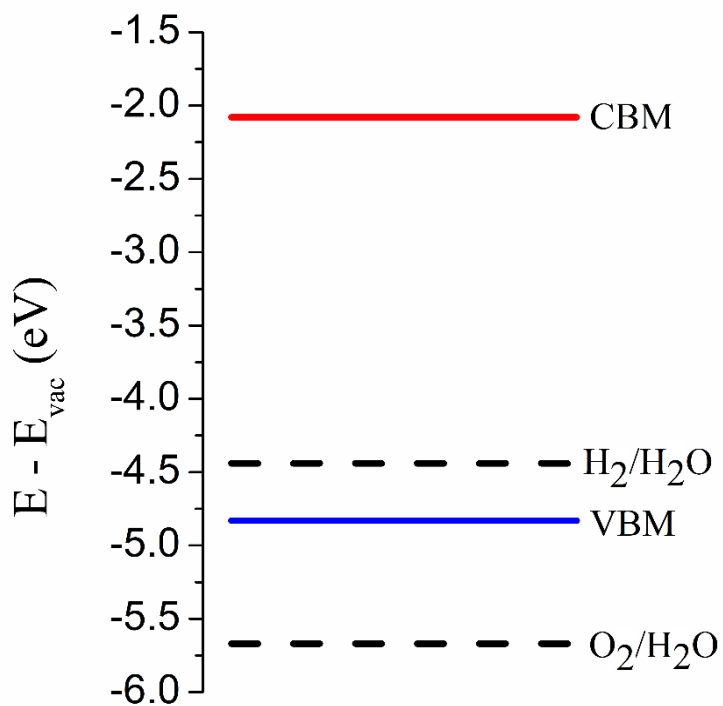

**Figure S12.** The calculated VBM and CBM positions of 2D $Zn_2VN_3$ with respect to the water redox potentials.